\documentclass[12pt]{iopart}
\usepackage{iopams}  
\usepackage{graphicx}  

\begin{document}
\title[EUV lines from W ions]
{EUV spectra of highly-charged 
ions W$^{54+}$-W$^{63+}$ relevant to ITER diagnostics}

\author{Yu~Ralchenko$^1$, IN~Draganic$^1$, JN~Tan$^1$, 
JD~Gillaspy$^1$, JM~Pomeroy$^1$, J~Reader$^1$, U~Feldman$^2$ 
and GE~Holland$^3$}

\address{$^1$ National Institute of Standards and Technology, 
Gaithersburg, MD 20899-8422, USA}
\address{$^2$ Artep Inc., Ellicott City, MD 21042, USA and
Space Science Division, Naval Research Laboratory, Washington, 
DC 20375-5320, USA}
\address{$^3$ SFA Inc., 2200 Defense Highway, Suite 405, Crofton MD 21114,
USA}
\ead{yuri.ralchenko@nist.gov}

\begin{abstract}

We report the first measurements and detailed analysis of extreme ultraviolet
(EUV) spectra (4 nm to 20 nm) of highly-charged tungsten ions W$^{54+}$ to
W$^{63+}$ obtained with an electron beam ion trap (EBIT).
Collisional-radiative modelling is used to identify strong electric-dipole 
and magnetic-dipole transitions in all ionization stages. These lines can be
used for impurity transport studies and temperature diagnostics in fusion
reactors, such as ITER. Identifications of prominent lines from several W ions
were confirmed by measurement of isoelectronic EUV spectra of Hf, Ta, and Au.
We also discuss the importance of charge exchange recombination for correct
description of ionization balance in the EBIT plasma.

\end{abstract}
\pacs{32.30.Rj, 32.70.Fw, 31.15.Am, 52.50.Hv}
\submitto{\jpb}
\maketitle

Future fusion reactors, such as ITER, are to reach temperatures of about 20
keV to 25 keV \cite{ITER}. While light elements (e.g., D, T, and C) will be
completely ionized in the plasma core interior, heavy impurity ions will still
possess a number of electrons. This could result in strong line emission,
primarily in the x-ray region. Such radiation is the source of power losses
that represent one of the major concerns for sustainable fusion. On the other
hand, high resolution x-ray spectroscopy of impurities offers reliable
measurements of important plasma parameters such as ion temperature $T_i$,
rotation velocities in toroidal and poloidal directions, and electron
temperature $T_e$ \cite{Donne}. Among all possible impurities in ITER,
tungsten is expected to be the most abundant, since according to current
planning the front surface of the divertor will be made of this element.

Extreme ultraviolet (EUV) line emission of impurities has attracted less
attention than the x-ray lines, in part because the flux of emitted photons in
the EUV region is typically smaller due to lower transition probabilities.
Feldman \etal \cite{Feldman} recently pointed out that a number of EUV lines
from highly-charged tungsten (ion charge $z$~$\gtrsim$~53) may be reliably
recorded in the ITER plasma and used for diagnostics of plasma temperature and
impurities transport. They proposed, in addition to the traditional grazing
incidence spectrometer, a system of segmented multilayer-coated telescopes for
registration of EUV lines, similar to what is used in solar corona diagnostics
(see, e.g., \cite{Soho}). Other possibilities for registration of EUV lines in
ITER are also discussed in the literature \cite{Donne}.

The objective of the present work is to investigate the EUV spectra from those
highly-charged W ions that will be abundant in the plasma of ITER. The
measurements of the spectra were performed with the Electron Beam Ion Trap
(EBIT) at the National Institute of Standards and Technology (NIST). EBIT is a
versatile light source, capable of producing nearly any ion charge state of
nearly any element. A fine control of the charge state distribution in the
trap is due to a very narrow electron energy distribution function (EEDF) of
the beam (width $\lesssim$ 60~eV) \cite{Width}. A detailed description of the
NIST EBIT can be found elsewhere \cite{EBIT}. The ions under study were
injected into the EBIT using a multi-cathode, metal vapor vacuum arc (MEVVA)
\cite{MEVVA} designed for a rapid change of the injected element without
downtime or alteration of experimental conditions. The EUV spectra between 4
nm and 20 nm were recorded with a grazing-incidence spectrometer described in
detail in \cite{Blagojevic}. The instrument's resolution is about 350,
corresponding to a resolving limit of about 0.03 nm. The W spectra were
calibrated with lines of highly-charged ions of Fe, with wavelengths taken
from the NIST Atomic Spectra Database \cite{ASD}.

\begin{figure}
\centering
\includegraphics[scale=0.90]{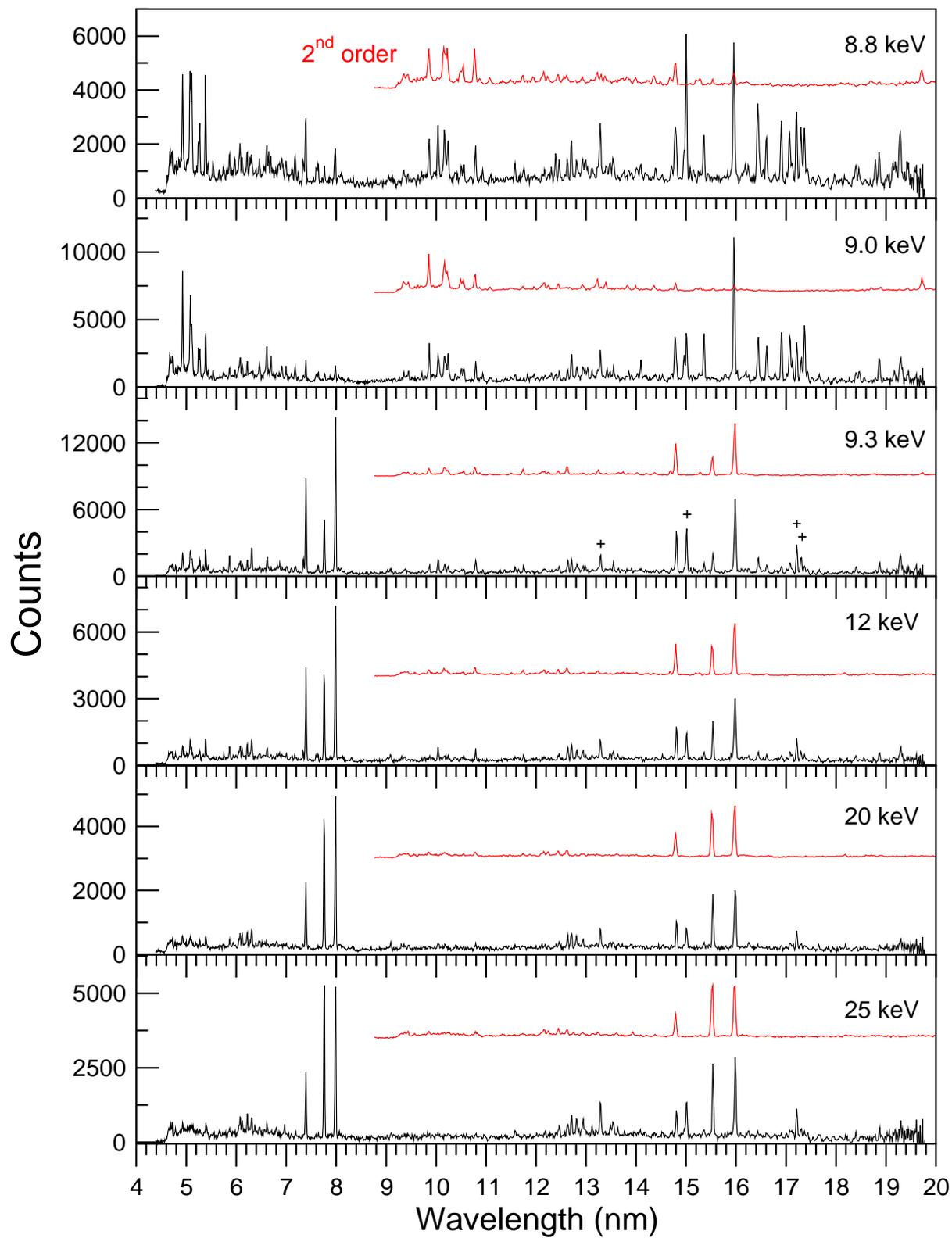}
\caption{\label{exper} (colour online) Experimental spectra at six beam energies
from 8.8 keV to 25 keV. For better identification, the thin red line shows the
second order spectrum. Oxygen impurity lines are marked by crosses.}
\end{figure}

The measured spectra for six nominal electron beam energies $E_B$ varying from
8.8~keV to 25~keV are presented in figure \ref{exper}. A typical value of the
beam current was about 150 mA.  Unlike our recent work on low-energy EUV
tungsten spectra \cite{EUV}, here we did not use a zirconium foil to filter
out light from wavelengths above 25 nm; this resulted in significantly higher
signal-to-noise ratio near the edges of the spectral range. For easier visual
identification of the second-order lines, the shifted spectra (red line) show
the same experimental spectra with wavelengths multiplied by 2. The spectra
also show a number of impurity lines, mainly from oxygen. The oxygen lines at
13.3 nm, 15.0 nm,  17.2 nm, and 17.3 nm are marked by crosses in the spectrum
for $E_B$ = 9.3 keV (figure \ref{exper}).

As with our previous studies of x-ray and EUV spectra of tungsten ions with
lower charges ($z$~=~37 to 47) \cite{EUV,XRAY}, identification of the spectral
lines relies upon collisional-radiative modelling performed with the
non-Maxwellian code NOMAD \cite{NOMAD}. The Flexible Atomic Code (FAC)
\cite{FAC}, based on the relativistic model-potential method and jj-coupling
scheme, was used to generate all relevant atomic data such as energy levels,
wavelengths, transitions probabilities, and collisional cross sections for the
ions W$^{54+}$ through W$^{65+}$. Previous comparisons with the EBIT tungsten
spectra at lower beam energies \cite{EUV,XRAY} showed that FAC provides
sufficiently accurate results for highly-charged ions. Using the FAC data,
NOMAD calculates the ionization balance, level populations, and line
intensities for the EBIT conditions. The EEDF in our simulations was modelled
as a Gaussian peak with a full width at half maximum of 60 eV, and the
electron density was set at $N_e$ = 2$\times$10$^{11}$~cm$^{-3}$. The
calculated spectra were convolved with a Gaussian instrumental function and
relative efficiency curve of the spectrometer \cite{Blagojevic}. 

The identifications of the spectral lines, the measured and calculated
wavelengths, transition types, and calculated radiative decay rates are
presented in table \ref{Table1}. The total uncertainties in the wavelengths
reported in the table are typically dominated by the uncertainty in the
spectrometer calibration, rounded up to one significant digit. An additional
statistical uncertainty due to fitting the reported line centers has been
added in quadrature. To facilitate classification, for each state we also
report a calculated level number in the energy-ordered list of levels within
the corresponding ion stage. All identified spectral lines correspond to
$\Delta n$ = 0 transitions within the $n$ = 3 shell. The $LS$-coupling
identifications of the upper and lower levels that were obtained using the
Cowan code \cite{Cowan} are only approximate due to strong spin-orbit
interaction.

\begin{table}
\caption{\label{Table1}Identified lines of highly-charged ions of tungsten. 
The experimental uncertainties for wavelengths are given in parentheses. 
The level number within the corresponding ion stage is shown in square
brackets. 
}
\begin{tabular}{ccrrcrrl}
\br
  & Isoelectronic & & & 
& $\lambda _{exp}$ & $\lambda_{th}$  & A \\
Ion     & sequence   & Lower level  & Upper level   &  Type  &   (nm)  &
     (nm)        &   (s$^{-1}$)\\
\mr
W$^{54+}$ & Ca & 3d$^2$ $^3$F$_2$ [1] & 3d$^2$ $^3$F$_3$ [3] 
& M1 & 17.078(3)  & 17.149 & 3.86$\times$10$^6$\\
          &      & 3d$^2$ $^3$F$_2$ [1] & 3d$^2$ $^3$P$_2$ [4] 
& M1 & 14.961(3)  & 14.980 & 1.81$\times$10$^6$\\
\\
W$^{55+}$ & K & 3p$^6$3d $^2$D$_{3/2}$ [1] & 3p$^6$3d $^2$D$_{5/2}$ [2] 
& M1 & 15.961(3)  & 16.007 & 2.60$\times$10$^6$ \\
          &     & 3p$^6$3d $^2$D$_{5/2}$ [2] & 3p$^5$3d$^2$ $^2$G$_{7/2}$ [6] 
& E1 & 6.622(3)   & 6.5873 & 5.65$\times$10$^6$\\
\\
W$^{56+}$ & Ar & 3p$^5$3d $^3$F$_{3}$ [4] & 3p$^5$3d $^3$F$_{4}$ [6] 
& M1 & 17.372(3)  & 17.453 & 2.09$\times$10$^6$ \\
          &      & 3p$^5$3d $^3$D$_{2}$ [5] & 3p$^5$3d $^1$D$_{2}$ [7] 
& M1 & 17.137(3)  & 17.189 & 8.71$\times$10$^5$ \\
          &      & 3p$^5$3d $^3$P$_{1}$ [3] & 3p$^5$3d $^1$D$_{2}$ [7] 
& M1 & 15.359(3)  & 15.358 & 1.85$\times$10$^6$ \\
          &      & 3p$^5$3d $^3$D$_{2}$ [5] & 3p$^5$3d $^3$D$_{3}$ [8] 
& M1 & 14.785(3)  & 14.787 & 2.54$\times$10$^6$ \\
          &      & 3p$^5$3d $^3$F$_{3}$ [4] & 3p$^5$3d $^3$D$_{3}$ [8] 
& M1 & 14.098(3)  & 14.087 & 7.79$\times$10$^5$ \\
          &      & 3p$^6$ $^1$S$_{0}$ [1]   & 3p$^5$3d $^3$P$_{1}$ [3] 
& E1 & 4.931(2)   & 4.9279 & 6.14$\times$10$^9$ \\
\\
W$^{57+}$ & Cl & 3p$^4$3d $^2$F$_{7/2}$ [5] & 3p$^4$3d $^4$D$_{7/2}$ [7] 
& M1 & 17.428(3)  & 17.530 & 6.07$\times$10$^5$\\
          &      & 3p$^4$3d $^2$F$_{7/2}$ [5] & 3p$^4$3d $^4$F$_{9/2}$ [8] 
& M1 & 16.911(3)  & 16.996 & 2.27$\times$10$^6$ \\
          &      & 3p$^4$3d $^4$D$_{5/2}$ [3] & 3p$^4$3d $^4$D$_{7/2}$ [7] 
& M1 & 16.613(3)  & 16.669 & 1.84$\times$10$^6$\\
          &      & 3p$^5$   $^2$P$_{3/2}$ [1] & 3p$^4$3d $^4$D$_{3/2}$ [2] 
& E1 & 5.254(2)   & 5.2558 & 2.60$\times$10$^9$ \\
          &      & 3p$^5$   $^2$P$_{3/2}$ [1] & 3p$^4$3d $^4$D$_{5/2}$ [3] 
& E1 & 5.121(2)   & 5.1209 & 4.20$\times$10$^9$ \\
\\
W$^{58+}$ & S & 3p$^3$3d $^3$F$_{3}$ [5] & 3p$^3$3d $^3$F$_{4}$ [7] 
& M1 & 16.444(3)  & 16.466 & 2.59$\times$10$^6$ \\
          &     & 3p$^4$  $^3$P$_{2}$  [1] & 3p$^3$3d $^3$D$_{2}$ [3] 
& E1 & 5.280(2)   & 5.2841 & 8.73$\times$10$^9$ \\
          &     & 3p$^4$  $^3$P$_{2}$  [1] & 3p$^3$3d $^3$F$_{3}$ [5] 
& E1 & 5.086(2)   & 5.0912 & 1.53$\times$10$^9$ \\
\\
W$^{59+}$ & P & 3s$^2$3p$^2$3d $^4$F$_{3/2}$ [2] & 3s$^2$3p$^3$ $^2$D$_{5/2}$ [5] 
& E1 & 10.042(3)  & 10.016 & 6.24$\times$10$^7$ \\
          &     & 3s$^2$3p$^3$ $^2$D$_{5/2}$ [5] & 3s3p$^4$  $^4$P$_{5/2}$     [7] 
& E1 & 7.607(3)   & 7.6260 & 9.81$\times$10$^9$  \\
          &     & 3s$^2$3p$^3$ $^2$P$_{3/2}$ [1] & 3s$^2$3p$^2$3d $^4$F$_{3/2}$ [2] 
& E1 & 5.396(2)   & 5.4037 & 1.24$\times$10$^{10}$ \\
\\
W$^{60+}$ & Si & 3s$^2$3p$^2$ $^1$D$_{2}$  [3] & 3s3p$^3$  $^3$P$_{2}$   [4] 
& E1 & 7.648(3)   & 7.6660 & 1.23$\times$10$^{10}$ \\
          &      & 3s$^2$3p$^2$ $^3$P$_{1}$  [2] & 3s3p$^3$  $^3$P$_{2}$   [4] 
& E1 & 7.189(3)   & 7.2436 & 1.54$\times$10$^{10}$ \\
\\
W$^{61+}$ & Al & 3s$^2$3p $^2$P$_{1/2}$ [1] & 3s3p$^2$ $^4$P$_{1/2}$ [2] 
& E1 & 7.404(2)   & 7.3953 & 3.88$\times$10$^{10}$ \\
          &      & 3s3p$^2$ $^4$P$_{1/2}$ [2] & 3s$^2$3p $^2$P$_{3/2}$ [3] 
& E1 & 6.318(3)   & 6.3246 & 3.05$\times$10$^8$  \\
\\
W$^{62+}$ & Mg & 3s$^2$ $^1$S$_{0}$ [1] & 3s3p $^3$P$_{1}$ [3] 
& E1 & 7.991(2)   & 7.9904 & 1.82$\times$10$^{10}$ \\
\\
W$^{63+}$ & Na & 3s $^2$S$_{1/2}$ [1] & 3p $^2$P$_{1/2}$ [2] 
& E1 & 7.769(2)   & 7.7617 & 4.37$\times$10$^{10}$ \\
\br
\end{tabular}
\end{table}

A straightforward method to match experimental and calculated line intensities
consists in varying the theoretical beam energy until a reasonable agreement
is achieved. While the calculated wavelengths agree well with the
measurements, our modelling required significantly lower beam energies to
reproduce the observed relative line intensities. For instance, for the
highest nominal beam energy of E$_B$ = 25 keV, the best-fit theoretical energy
was only 10.5 keV, while for E$_B$ = 8.8 keV the fit energy was about 6.2 keV.
Although space-charge effects in the trap may effectively reduce the beam
energy, for the present values of the nominal voltage and beam current this
would result in a less than 300 eV shift. 

Calculation of ionization balance in an EBIT plasma was a subject of several
studies (e.g., \cite{Penetrante,Liu,Lu,Kalagin} and references therein). Each
ion is normally modelled as one state with no internal structure, and
approximate formulas, such as Lotz formula for ionization cross sections, are
used for description of atomic processes. In constrast, our
collisional-radiative modelling emphasizes the atomic physics component of
simulations by implementing  detailed descriptions of charge stages (about
1000 levels per ion) and relevant physical processes affecting populations of
atomic levels. This is a necessary condition for the accurate modelling of
well-resolved EBIT spectra.

It is known (e.g., \cite{EUV}) that the relative line intensities {\it within}
a particular ionization stage are only weakly sensitive to variations in the
EBIT beam energy. Therefore the observed disagreement with the simulations is
most likely due to modification of the ionization distribution in the EBIT
plasma. This discrepancy can be explained by charge exchange (CX) with neutral
particles (mostly nitrogen and oxygen) that are always present in the trap.
First, note that the only important electron-ion recombination process in the
EBIT is the radiative recombination. Dielectronic recombination (DR) would
generally be very important for the high ion charges in question; however, it
is an energy-selective process, and for all cases considered here the electron
beam energy is outside the range of importance for DR. Our present
calculations show that the typical radiative recombination cross sections for
the relevant ions of tungsten are on the order of 10$^{-21}$  cm$^2$ (see also
\cite{KimP}). It is well known that the CX cross sections are normally much
larger. The Classical Trajectory Monte-Carlo (CTMC) theory \cite{Olson} gives
the approximate dependence as:

\begin{equation}
\label{sigcx}
\sigma_{CX} \approx z \times 10^{-15} ~cm^2,
\end{equation}
and thus for 60 to 70 times ionized atoms the CX cross section is close to
10$^{-13}$ cm$^2$.  This simple formula agrees within 40\% with the
semiempirical scaling law for charge exchange \cite{Liu,Selberg} that is
often used in charge distribution simulations of EBIT plasmas.

\begin{figure}
\centering
\includegraphics[scale=0.65]{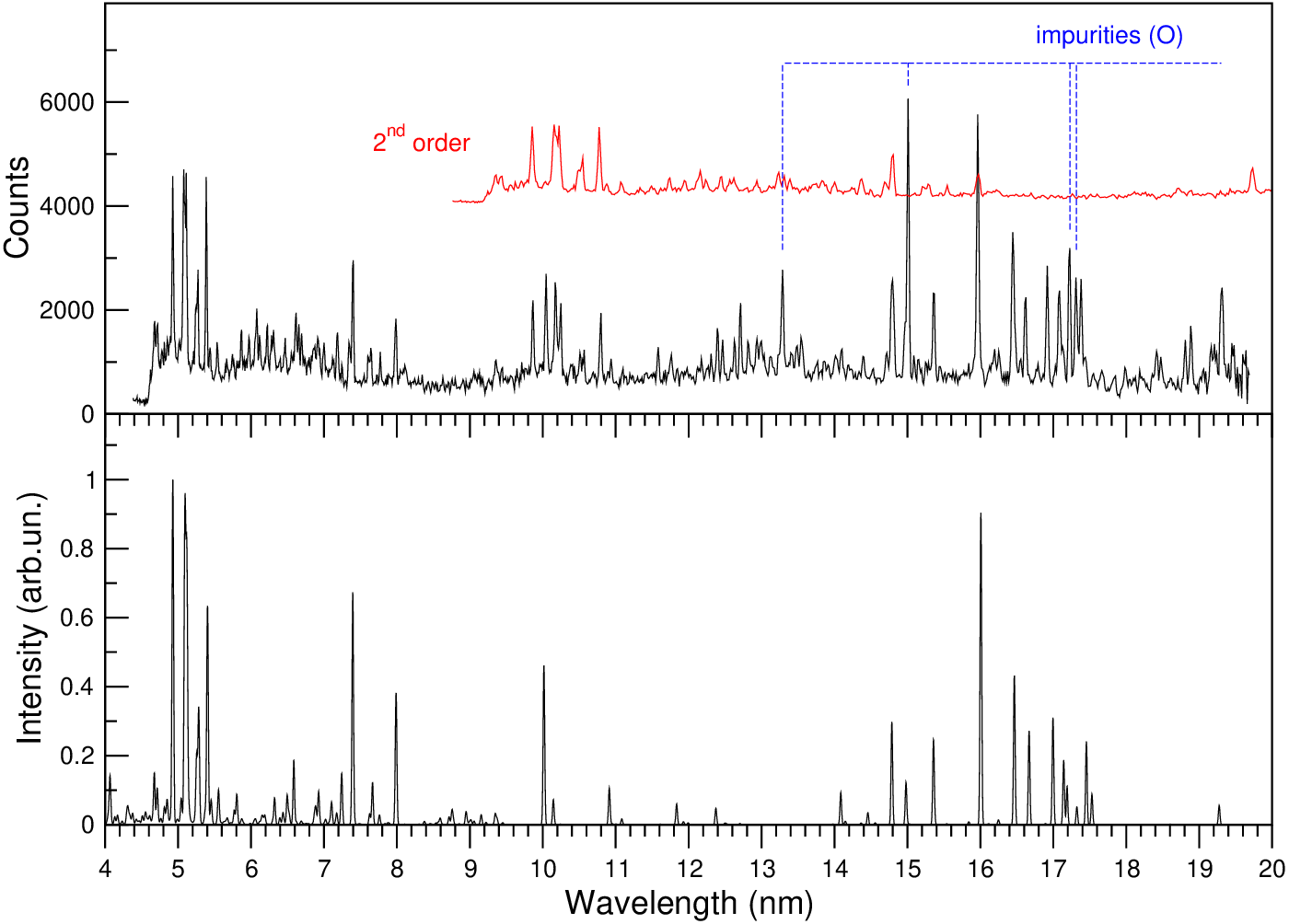}
\caption{\label{sim}(colour online) Comparison of experimental (top) and theoretical (bottom) 
spectra at the beam energy $E_B$ = 8.8 keV. The second-order spectrum is shown
by the red line. The positions of the strongest impurity lines from
highly-ionized oxygen are indicated by dashed lines.}
\end{figure}

The CX rate depends on the density of the neutrals $N_n$, relative velocity of
neutrals and tungsten ions $v_r$, and the cross section $\sigma_{CX}$. For
simplicity, the product $\langle \sigma_{CX} v_r \rangle$, averaged over the
relative energy distribution of neutrals and W ions, can be replaced by the
product of $\sigma_{CX}$ and effective averaged velocity $\tilde{v}_r$, so
that the CX rate becomes  $R_{CX} = N_n \sigma_{CX} \tilde{v}_r$. Both $N_n$
and $\tilde{v}_r$ are unknown parameters that may actually vary from
measurement to measurement. (A slight enhancement of lower ionization stages
at $E_B$ = 9.0 keV vs. $E_B$ = 8.8 keV may be due to this uncertainty.)
However, since they enter only as a product, one can use this single parameter
to fit the ionization balance. As there exist no calculations of CX for such
highly-charged ions, we used the CX cross section of Eq. (\ref{sigcx}) and
varied the $N_n \tilde{v}_r$ product to reach agreement with experimental
spectra. The example of correspondence for the most complex spectrum at $E_B$
= 8.8 keV is presented in Fig. \ref{sim}. One can see that the calculated
spectrum that uses only one fit parameter (the theoretical beam energy was
fixed at 8.8 keV) correctly reproduces relative intensities for practically
all observed lines. The derived value of $N_n  \tilde{v}_r$ $\approx$
4$\times$10$^{13}$  cm$^{-2}$s$^{-1}$ is consistent with general estimates for
$N_n$ and $\tilde{v}_r$ expected inside the EBIT. The best theoretical fit for
the experimental spectrum at $E_B$ = 25 keV is obtained with $N_n 
\tilde{v}_r$ $\approx$ 1$\times$10$^{13}$  cm$^{-2}$s$^{-1}$.

As mentioned above, our simulations were performed at $N_e$ =
2$\times$10$^{11}$ cm$^{-3}$. Without CX, the ionization balance is
practically independent of $N_e$ for a several-orders-of-magnitude span of
densities. With CX, variations in $N_e$ would certainly modify ionization
rates. However, since $N_n \tilde{v}_r$ is a free parameter in our
simulations, it can be appropriately adjusted to restore agreement with the
experimental spectra. Thus, dependence on $N_e$ is effectively removed from
the modelling.

At the lowest energies $E_B$ = 8.8 keV and 9.0 keV (see figure \ref{exper}),
one can clearly identify two groups of strong lines, namely, in the range of 4
nm to 8 nm and 14.5 nm to 17.5 nm. These lines originate from ions W$^{54+}$
through W$^{60+}$. As the beam energy increases, the ionization distribution
shifts to higher charge states, and  the three lines from Al-like W$^{61+}$
($\lambda$~=~7.404~nm), Mg-like W$^{62+}$ ($\lambda$~=~7.991~nm), and Na-like
W$^{63+}$ ($\lambda$~=~7.769~nm) ions become the most prominent. The variation
of their intensities agrees with the changes in the beam energy. These three
lines will be easily resolved in plasmas of fusion reactors, and in addition,
they cover a range of only 0.6 nm, so that determination of their relative
intensities will be less dependent on the spectral sensitivity of EUV
spectrometers. 

\begin{figure}
\centering
\includegraphics[scale=0.70]{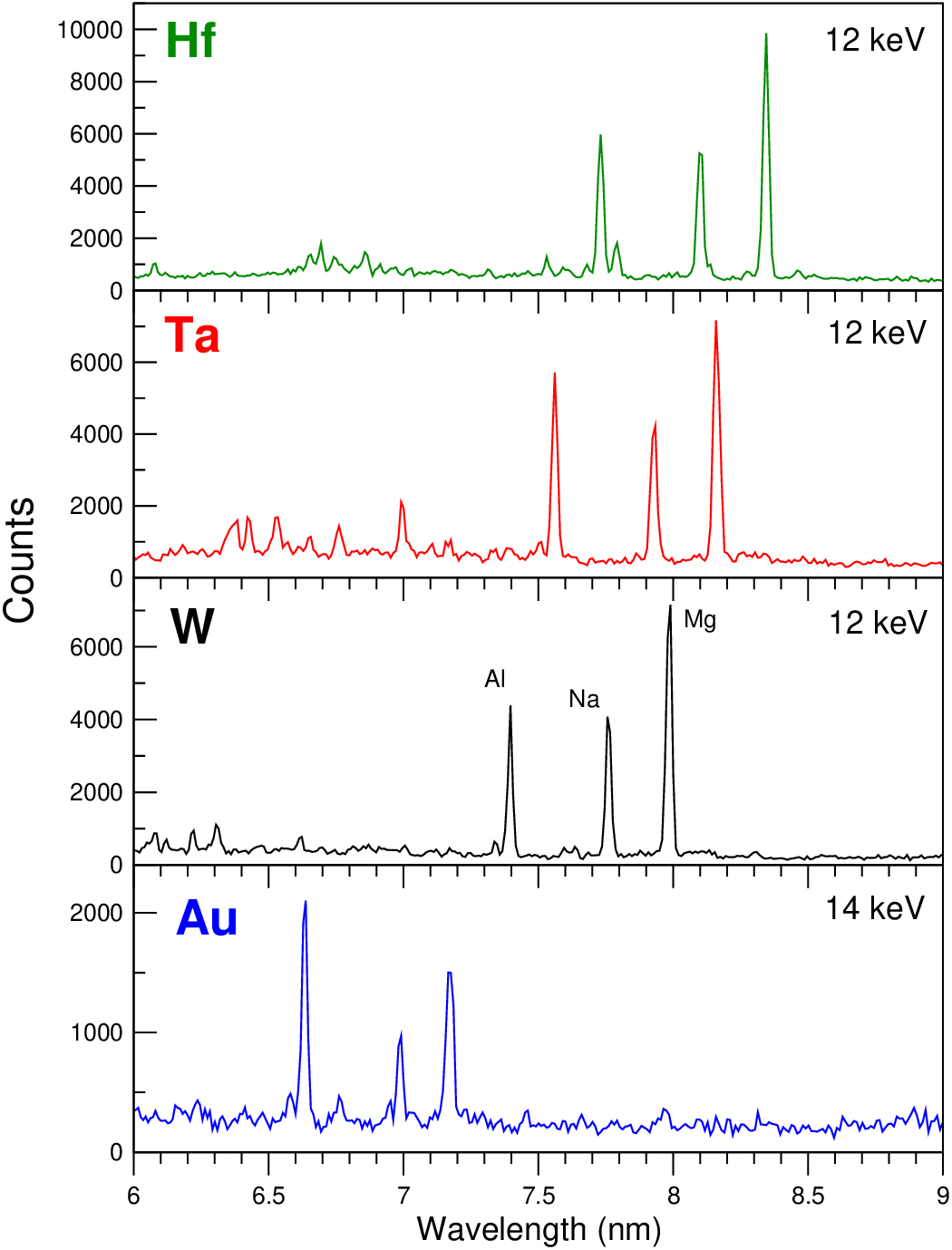}
\caption{\label{HfAu}(colour online) Experimental spectra for Hf (beam energy 12 keV), 
Ta (12 keV), W (12
keV), and Au (14 keV). The three strongest lines originate from (short to
long) Al-, Na-, and
Mg-like ions, respectively.}
\end{figure}

\begin{figure} 
\centering
\includegraphics[scale=0.50]{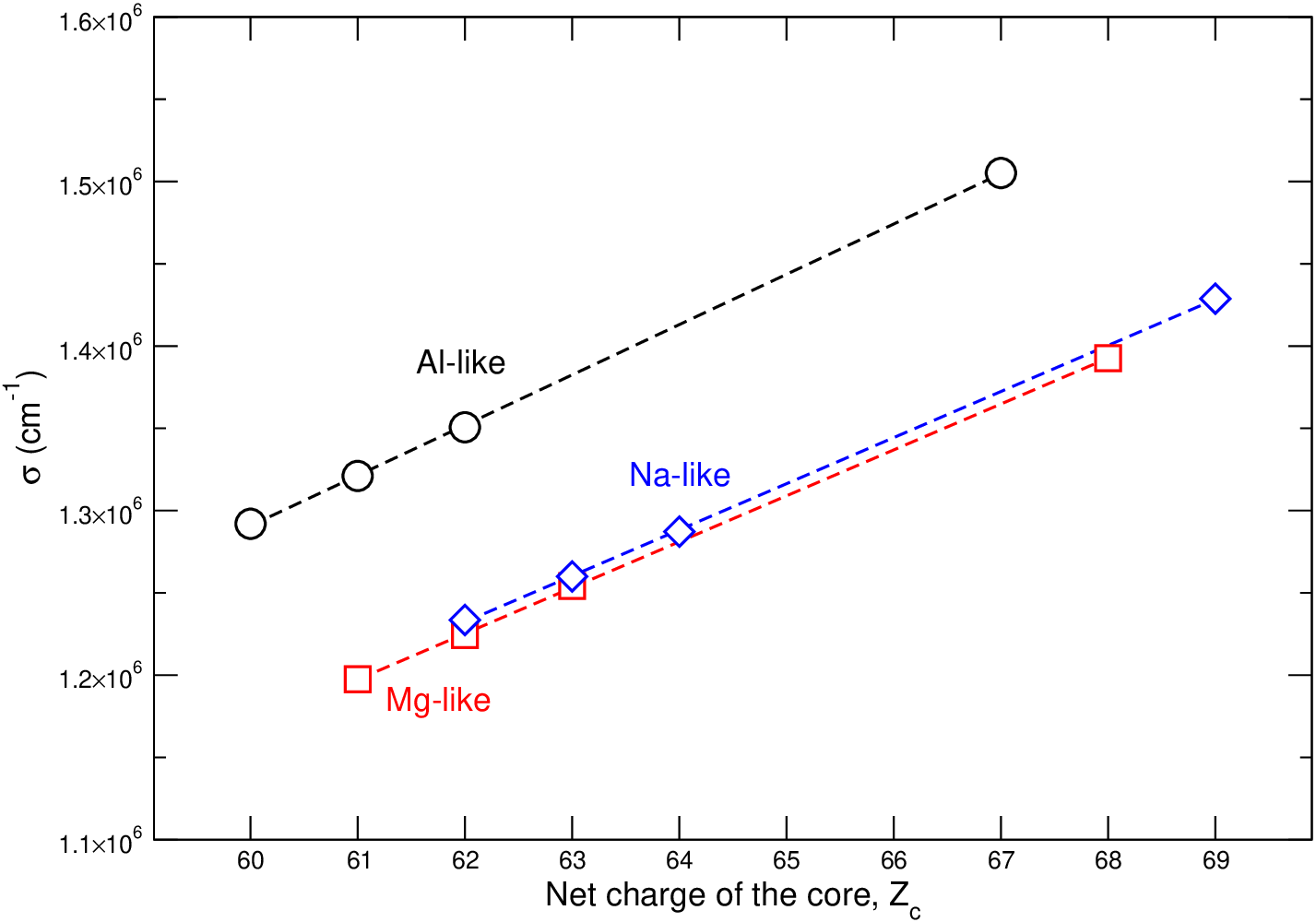}
\caption{\label{fit}(colour online) $Z_{c}$-dependence of wavenumbers (in cm$^{-1}$) for the
three strongest lines of Fig. \ref{HfAu}: Al-like -- circles, Mg-like --
squares, Na-like -- diamonds. The linear fits are shown by dashed lines. The
error bars for wavenumbers are completely within the symbols.} 
\end{figure}

To confirm our identification of these strong lines in W, we also measured the
EUV spectra for Hf ($Z$=72), Ta ($Z$=73), and Au ($Z$=79). The
nominal beam energy for Au was 14~keV, while the Hf and Ta spectra were
recorded at $E_B$ = 12~keV. This selection of energies provided ionization
distributions similar to the 12~keV measurements for W. Figure \ref{HfAu}
shows the recorded spectra between 6~nm and 9~nm, including the W results. The
measured and calculated wavelengths for Hf, Ta, and Au are presented in Table
\ref{Table2}. As these lines correspond to the $\Delta n = 0$ transitions, the
leading term in the transition energy difference should be proportional to the
net charge of the atomic core $Z_{c}$ = $z + 1$. This dependence is well confirmed in
Fig. \ref{fit} which shows the linear fit results for wavenumbers $\sigma$. 
The fit $\sigma$~[cm$^{-1}$] = A $\cdot$ $Z_{c}$ -- B agrees with the
experimental data to within 0.1~\% for the following fit coefficients (A,B):
(3.056$\times$10$^4$, 5.428$\times$10$^5$) for the Al-like line,
(2.791$\times$10$^4$, 5.052$\times$10$^5$) for the Mg-like line, and
(2.801$\times$10$^4$, 5.043$\times$10$^5$) for the Na-like line. The derived
coefficients can be used to predict wavelengths of these strong lines in other
elements.

\begin{table}
\caption{\label{Table2}Measured and calculated wavelengths (in nm) of the 
strongest lines from Al-, Mg-, and
Na-like ions of Hf, Ta, and Au. The experimental wavelength uncertainty is 
$\pm$ 0.002 nm for all lines in this table.}
\begin{indented}
\item[]
\begin{tabular}{crclcccccc}
\br
&&&&\centre{2}{Hf}&\centre{2}{Ta}&\centre{2}{Au}\\
\ns
&&&&\crule{2}&\crule{2}&\crule{2}\\
Sequence&\centre{3}{Transition}&expt&calc&expt&calc&expt&calc\\
\mr
Al& 3s$^2$3p $^2P_{1/2}$ & -- &3s3p$^2$ $^4P_{1/2}$ &7.741 &7.7297 & 7.570 
&7.5598 & 6.644 &6.6459 \\
Mg& 3s$^2$ $^1S_0$ & -- &3s3p $^3P_1$ &8.352 &8.3456 & 8.169 &8.1652 & 7.181 &7.1902 \\
Na& 3s $^2S_{1/2}$ & -- &3p $^2P_{1/2}$ &8.107 &8.0973 & 7.937 &7.9270 & 6.999 &7.0030 \\
\br
\end{tabular}
\end{indented}
\end{table}

There exist only a few theoretical works on calculated energy levels,
wavelengths, and transition probabilities for W ions with 54 $\leq$ $z$ 
$\leq$ 63, with most efforts related to Na-like W$^{63+}$. Ali and Kim
\cite{Ali} used multiconfiguration Dirac-Fock (MCDF) method to calculate the
wavelength of the $3s^63d$~$^2D_{3/2}$ -- $3s^63d$~$^2D_{5/2}$ transition in
W$^{55+}$, and their value of $\lambda$ = 15.959 nm is within the experimental
uncertainty. The wavelength of the 3s$^2$ $^1S_0$ - 3s3p $^3P_1$ transition in
Mg-like W$^{62+}$ was calculated at $\lambda$ = 8.0509 nm using the
relativistic model-potential method \cite{Ivanova}, compared to our measured
value of 7.991(2) nm. Among the recent calculations for the 3s$_{1/2}$ -
3p$_{1/2}$ transition in the Na-like ion, the MCDF \cite{Seely,Baik} and
many-body perturbation theory (MBPT) \cite{Johnson} results should be
mentioned. The latter give a wavelength $\lambda$ = 7.7659 nm in very good
agreement with the present experiment, and the MBPT transition probability A =
4.296 $\times$ 10$^{10}$ s$^{-1}$ agrees with our calculations within 2~\%.
The ab initio MCDF wavelength of Seely \etal \cite{Seely} $\lambda$ = 7.7479
nm  is also close to the measured value, while Baik \etal's  result of
$\lambda$ = 7.8184 nm \cite{Baik} is higher by 0.05 nm. Finally, the
wavelength predicted in \cite{Seely} by extrapolation along isoelectronic
sequence, $\lambda$ =  7.7661 nm, agrees well with the experiment.

The low electron density in the EBIT, two to three orders of magnitude lower than
that in typical tokamaks, allows one to observe weak spectral lines that might
be collisionally quenched in fusion reactors. A recently discussed example is
provided by the magnetic-octupole line in the Ni-like W$^{46+}$ \cite{M3},
with a radiative transition probability on the order of $10^4$ s$^{-1}$, that
becomes quenched at densities above 10$^{12}$ cm$^{-3}$. The higher degree of
ionization for the lines considered here should be more favorable for their
survival as the most important $\Delta n$=0 collisional cross sections
decrease with an increase of ion charge. To test this assumption, we used the
NOMAD code to solve the steady-state system of rate equations for
approximately 6000 levels of the ions W$^{54+}$ through W$^{65+}$ at an
electron temperature $T_e$ = 10$^4$ eV and density $N_e$ = 10$^{14}$
cm$^{-3}$. It was found that for all lines in table \ref{Table1}, including
the weakest lines with transition probabilities A $\lesssim$ 10$^6$~s$^{-1}$,
the radiative decays are still the dominant depopulation process for the upper
levels of corresponding transitions. This result remains valid for a large
range of electron temperatures since the typical energy difference for
collisional processes is much smaller than $T_e$. Therefore, these lines will
not be collisionally quenched in the ITER plasma.

To conclude, we presented here the first measurements of EUV spectra from
highly-charged ($z$ = 54 to 63) ions of W that are important for plasma
diagnostics in ITER-class fusion reactors. Using advanced
collisional-radiative modelling of line spectra from the non-Maxwellian plasma
of EBIT, we identified 26 new electric-dipole and magnetic-dipole lines from
10 consecutive ionization stages W$^{54+}$ through W$^{63+}$. We also measured
wavelengths of three important spectral lines for Al-, Mg-, and Na-like ions
of Hf, Ta, and Au and used them to confirm our identifications of
corresponding lines in W. These prominent lines may be used as tracer
identifications in the ITER plasma interior. The increasing importance of
charge exchange for ionization balance of high-Z ions in the EBIT plasmas was
pointed out as well. A detailed description of the modelling and a discussion
of various spectral features will be provided elsewhere.

\ack

This research was performed while one of us (I.N.D.) held a National Research
Council Research Associateship Award at NIST-NIH. Our work was supported in
part by the Office of Fusion Energy Sciences of the U.~S. Department of
Energy. Valuable comments from an anonymous referee are highly appreciated.

{\it Note: After the manuscript was accepted for publication, we were informed
that three strong lines in Au were recently reported in \cite{Traebert}.}

\end{document}